# Generalized Bimode Equivalent Circuit of Arbitrary Planar Periodic Structures for Oblique Incidence

Fernando Conde-Pumpido, Gerardo Perez-Palomino, J. Ramón Montejo-Garai, and Juan E. Page

*Abstract*—This work presents, for the first time, a generalized bimode Foster's equivalent circuit for characterization of 2-D Planar Periodic Structures (PPSs) with arbitrary geometry at oblique incidence. It considers the interactions between the fundamental TE and TM modes without any restriction within the bimode bandwidth of the geometry. The proposed circuit is only composed of frequency-independent LC elements, which can be extracted systematically from electromagnetic (EM) simulations. The reactive immittances obtained in the process fulfill the Foster's theorem, enabling the design process of PPS-based devices using standardized synthesis techniques from circuit theory. To demonstrate its viability and general nature, equivalent circuits are extracted for different single- and multilayer PPS composed of rotated dipoles under oblique incidence θ=20º, φ=30º, and including dielectrics. Excellent agreement is found between the response of the circuit model and the EM simulation in all cases. Finally, to validate experimentally the proposed equivalent circuit and highlight its applicability, a 90º reflective Linear-Polarization (LP) Rotator centered at 25 GHz and under oblique incidence, θ=30º, φ=0º (TE), is designed, manufactured, and tested. The agreement between the circuit response, the EM simulation and the measurement underlines the potential of the new equivalent circuit for PPS design under oblique incidence.

## I. Introduction

NON-GUIDED communications are well established both within access and core networks. Space communications rely solely on them, making microwave links a critical piece of space operations. Being both spectrum and power scarce resources, devices that ensure the correct polarization, frequency, and radiation pattern of in- and outcoming signals are becoming ubiquitous [1]-[2]. In this context, single or multilayer planar periodic structures (PPSs), comprising a cascade of 2-D periodic printed metallic patterns stacked with dielectrics, are becoming a prevalent technology. They are suitable for designing polarization control devices for open-space applications in both transmission and reflection, such as linear-to-circular (LP-CP) polarization converters [3]-[5], or 90º linear polarization rotators [6]-[7]. They are also used to implement spatial and spectral filters, commonly known as frequency-selective surfaces (FSS) [8]-[11]; and aperture antennas (reflectarrays [12]-[14] and transmitarrays [15]-[16]).

The design process of these devices usually does not follow a systematical approach, and instead it is based on using EM optimizations that are computationally expensive, especially for multilayer devices. Most alternatives reduce this expensive step by building mathematical tools that can model, up to a degree of accuracy, the response of a range of parametric variations of a geometry. The accuracy obtained by them is good enough to be used as a predesign tool, lifting the heavier burden of the EM simulator for the main process, and leaving it only for final fine-tuning optimizations. Latest reports on these models include the use of trained artificial neural networks (ANN) [17] or support-vector machines (SVM) [18]. However, these tools usually require of complex training sets that must be repeated every time the structure (number of layers, shape) is changed.

Alternatively, the equivalent circuits are widely considered as a powerful tool to analyze and design PPS-based devices, since they allow taking advantage of physical concepts to improve the efficiency and accuracy of the model. Numerous equivalent circuits have been reported in the literature to model PPS-based devices, most of them phenomenological and purely reactive. They are usually obtained by means of non-standardized methods for the structure under study [19]-[23] or using systematic procedures to cover more geometries [24]-[27], although at the expense of formulations and approaches that are reduced to a more limited number of PPS and excitations. An extensive review on the state-of-the-art of these can be found in [28]-[30] together with their electrical limits. All these circuits usually lead to complex topologies comprising frequency-dependent elements, which turns them unsuitable as a design tool.

The Foster's circuits are a good alternative, as these encompass all the energetic interactions in lumped elements and therefore are the simplest in terms of number of elements.

Manuscript received XXX, XXXX This work was supported in part by the Spanish Government under Grant PID2020-116968RB-C33 (DEWICOM) and PID2020-114172RB-C22 (ENHANCE-5G) both funded by MCIN/AEI/10.13039/501100011033 (Agencia Estatal de Investigación). F. Conde-Pumpido, G. Perez-Palomino, J. Ramón Montejo-Garai, and Juan E. Page are with the Group of Applied Electromagnetics (GEA), Information Processing and Telecommunications Center, Universidad Politécnica de Madrid, E-28040, Madrid, Spain (e-mail: fernando.conde-pumpido.velasco@upm.es, gerardo.perezp@upm.es).

Digital Object Identifier 10.1109/TAP.2021.XXXXXXX

Moreover, these circuits can be represented with canonical topologies whose elements can be extracted systematically, thus enabling the application of design strategies from the circuit theory [31]. Some works have presented Foster's circuits that model PPSs within the non-diffractive band (or bimode) [32]-[34], and other variations to account for the effects of the conductive losses [35]. However, all of them are only suitable to model geometries where there is no interaction between the fundamental modes (or it can be considered negligible); this simplification occurs at normal incidence for geometries that exhibit symmetries with respect to the periods, and for certain geometries under an oblique incidence (mainly on the main planes).

The most challenging effect in a PPS to model using an equivalent circuit approach is the interaction between the fundamental modes. Strong interactions usually imply that the phenomenological circuits are so complex that, even if the topology can be found, they are often unsuitable in practice. A simple bimode Foster's equivalent circuit that appropriately models the behavior and the interaction between the fundamental TEM modes (Vertical and Horizontal) for any geometry has been presented recently [36]. It has proven to be valuable for designing polarization-control devices in a systematic way [4]. However, that circuit does not account for arbitrary angles of incidence, limiting its validity for designing devices that work in oblique incidence, or modelling the behavior of aperture antennas such as reflectarrays or transmitarrays, where each element is excited by a particular angle of incidence.

This work presents, for the first time, a bimode Foster's equivalent circuit able to model the electrical behavior within the bimode bandwidth of PPSs exhibiting arbitrary geometry under oblique incidence. In this way, all the gained knowledge in circuit theory is available, opening the field of applications by incorporating sophisticated synthesis techniques previously developed like the coupling matrix, the extracted pole technique and others to the PPS design. Thus a systematic synthesis process provides key information about the electrical limits of the devices to be designed in terms of canonical circuit topologies and distributed elements, as the minimum number of layers required to cope with the electrical specifications.

The paper is organized as follows. Section II presents the equivalent circuit and summarizes its theoretical bases under both, vacuum conditions and considering dielectrics. A brief discussion about the advantages and limitations of the proposed circuit is also disclosed. In Section III, numerical validation of the model is performed at the angle $\theta = 20º$, $\varphi = 30º$, comparing the results obtained from the equivalent circuit to those from a full-wave simulation in the case of a rotated dipole, both with and without dielectric layers. Another contribution in this section is the partial study of the rotated dipole from the point of view of a Foster's circuit at oblique incidence, showing maps of capacitances and inductances for a sweep of lengths and rotations. Finally, Section IV is devoted to designing a reflective polarization rotator using the equivalent circuit and double-L metallizations. The device transforms a TE mode impinging on a PPS at $\theta = 30º$ into a TM mode at 25 GHz. The design was manufactured and tested to check the validity of the proposed circuit and highlight its applicability

## II. FOSTER'S EQUIVALENT CIRCUIT FOR OBLIQUE INCIDENCE

In this section, the Foster's equivalent circuit for a PPS exhibiting arbitrary shape and excited at oblique incidence is derived. Firstly, the set of modes of the rectangular waveguide with periodic boundary conditions (PBC) on its walls is obtained, allowing to consider the scattering problem of a PPS from the modal analysis perspective. Secondly, a bimode equivalent circuit only made up of frequency- independent lumped elements is introduced. This circuit considers simultaneously the TE and TM fundamental modes and their interaction. Finally, a generalization that includes the introduction of dielectrics and multilayer structures is presented together with a discussion of the advantages, limitations and reliability of the proposed circuit.

### A. Theoretical Base

Let us consider a PPS with rectangular period ($P_x$ and $P_y$) placed in the vacuum on the plane $z = 0$. It is made up of a single layer and contains only infinitely thin, perfect-electrical-conductor (PEC) metallizations of arbitrary shape (Fig. 1a), which are periodically distributed across the plane forming a 2-D planar array of infinite elements or unit cells. Now, consider a plane wave travelling in the $\hat{n}$ direction defined by the spherical angles $\theta_i$ and $\varphi_i$ (Fig. 1a), which impinges on the PPS. If the structure is analyzed from the perspective of a guided transmission system with translational symmetry along $\hat{z}$ axis, two rectangular waveguides with PBC (labeled as $A$ and $B$) can be identified (Fig. 1b), each of which located at one side of the discontinuity. Since the excitation imposes a certain periodicity condition at the walls of the waveguide related to $\theta_i$ and $\varphi_i$, a complete orthogonal set of modes arise for each ($\theta_i$, $\varphi_i$) pair in both waveguides.

To calculate the modes of a rectangular waveguide with PBC, and given that the waveguide shows translational symmetry, the Helmholtz equation for the longitudinal components of the TE and TM modes must be solved [37]:

$$\Delta_t F_{TE,TM}(x,y) + k_c^2 F_{TE,TM}(x,y) = 0 \quad (1)$$

with the following boundary conditions [38]:

$$PBC \begin{cases} F_{TE,TM}(x+P_x,y) = \chi_x F_{TE,TM}(x,y) \\ F_{TE,TM}(x,y+P_y) = \chi_y F_{TE,TM}(x,y) \\ \hat{x} \cdot \nabla_t F_{TE,TM}(x+P_x,y) = \chi_x \hat{x} \cdot \nabla_t F_{TE,TM}(x,y) \\ \hat{y} \cdot \nabla_t F_{TE,TM}(x,y+P_y) = \chi_y \hat{y} \cdot \nabla_t F_{TE,TM}(x,y) \end{cases} \quad (2)$$

where:

$$\begin{cases} \chi_x = e^{-jk_0 P_x \hat{n}\cdot\hat{x}} = e^{-jk_0 P_x \sin\theta_i \cos\varphi_i} \\ \chi_y = e^{-jk_0 P_y \hat{n}\cdot\hat{y}} = e^{-jk_0 P_y \sin\theta_i \sin\varphi_i} \end{cases} \quad (3)$$

$$\hat{n} = \sin\theta_i \cos\varphi_i \cdot \hat{x} + \sin\theta_i \sin\varphi_i \cdot \hat{y} + \cos\theta_i \cdot \hat{z} \quad (4)$$

and $k_0$ is the vacuum phase constant.

If $\chi_x = \chi_y = 1$ (corresponding to the normal incidence case), the modes are TEM, which were already studied in [36] from a Foster's circuit point of view. In this work, the focus is on the condition $|\chi_x| = |\chi_y| = 1$, corresponding to the case of oblique incidence. In this case, the longitudinal and transversal components of the TE and TM modes are:

$$H_z = \sum_{m=-\infty}^{\infty}\sum_{n=-\infty}^{\infty} \underbrace{H_{mn} e^{-jk_{xm}x} e^{-jk_{yn}y}}_{F_{TE}(x,y)} e^{-jk_z z} \quad (TE) \quad (5)$$

$$E_z = \sum_{m=-\infty}^{\infty}\sum_{n=--\infty}^{\infty} \underbrace{E_{mn} e^{-jk_{xm}x} e^{-jk_{yn}y}}_{F_{TM}(x,y)} e^{-jk_z z} \quad (TM) \quad (6)$$

where:

$$k_{zmn} = \sqrt{k_0^2 - (k_{xm}^2 + k_{yn}^2)} = \sqrt{k_0^2 - k_{cmn}^2} \quad (7)$$

and

$$\begin{cases} k_{xm} = -\dfrac{\arg(\chi_x)}{P_x} = k_0 \sin\theta_i \cos\varphi_i + 2\pi m/P_x & m \in \mathbb{Z} \\ k_{yn} = -\dfrac{\arg(\chi_y)}{P_y} = k_0 \sin\theta_i \sin\varphi_i + 2\pi n/P_y & n \in \mathbb{Z} \end{cases} \quad (8)$$

The transversal components of the electromagnetic field for each mode $(m, n)$ can be written as [37]:

$$\vec{E}_t = -\frac{j\omega\mu}{k_{cmn}^2}\nabla_t H_z \times \hat{z}; \quad \vec{H}_t = -\frac{jk_z}{k_{cmn}^2}\nabla_t H_z \quad (TE) \quad (9)$$

$$\vec{E}_t = -\frac{jk_z}{k_{cmn}^2}\nabla_t E_z; \quad \vec{H}_t = \frac{j\omega\varepsilon}{k_{cmn}^2}\nabla_t E_z \times \hat{z} \quad (TM) \quad (10)$$

and the modal impedances for each mode are:

$$Z_{TE} = \frac{\hat{z}\times\vec{E}_t}{\vec{H}_t} = \frac{\omega\mu_0}{k_{zmn}} \quad (11)$$

$$Z_{TM} = \frac{\hat{z}\times\vec{E}_t}{\vec{H}_t} = \frac{k_{zmn}}{\omega\varepsilon_0} \quad (12)$$

The above expressions are also known as *Floquet modes*, although they have been obtained solving the rectangular waveguide with PBC [38], instead of solving a free space problem [39].

From (7), it can be deduced that there are two fundamental modes ($m = n = 0$ for TE and TM), which are degenerated and exhibit zero cutoff frequency, since $k_{z00} = k_{z0} \geq 0$ for any value of $(\theta_i, \varphi_i)$ and frequency. From the above equations, the expressions of the fundamental TE and TM modes are obtained:

$$TE \begin{cases} \vec{E}_{TE} = \vec{E}_{tTE} = \hat{u}_\perp \cdot a_{TE} e^{-jk_0\hat{n}\cdot\vec{r}} \\ \vec{H}_{TE} = \vec{H}_{tTE} + H_{zTE}\hat{z} = \hat{u}_\parallel \cdot (a_{TE}/\eta_0) e^{-jk_0\hat{n}\cdot\vec{r}} \end{cases} \quad (13)$$

$$TM \begin{cases} \vec{E}_{TM} = \vec{E}_{tTM} + E_{zTM}\hat{z} = \hat{u}_\parallel \cdot a_{TM} e^{-jk_0\hat{n}\cdot\vec{r}} \\ \vec{H}_{TM} = \vec{H}_{tTM} = -\hat{u}_\perp \cdot (a_{TM}/\eta_0) e^{-jk_0\hat{n}\cdot\vec{r}} \end{cases} \quad (14)$$

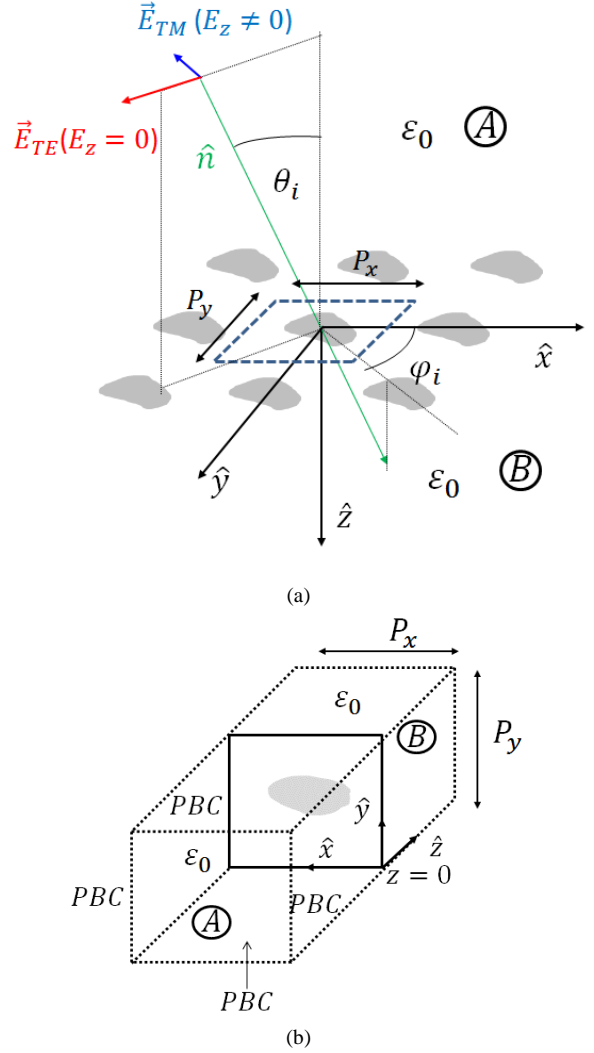

Fig. 1. (a) Monochromatic plane wave travelling in $\hat{n}$ direction and impinging upon a PPS at oblique incidence (b) Rectangular waveguides with Periodic Boundary Conditions (PBC).

The unit vectors $\hat{u}_\parallel$ and $u_\perp$ are ($\hat{r}$ is the position vector):

$$\begin{aligned} \hat{u}_\perp &= -\sin\varphi_i \cdot \hat{x} + \cos\varphi_i \cdot \hat{y} \\ \hat{u}_\parallel &= -\cos\theta_i(\cos\varphi_i \cdot \hat{x} + \sin\varphi_i \cdot \hat{y}) + \sin\theta_i \cdot \hat{z} \end{aligned} \quad (15)$$

and define the directions of the electric field in TE and TM modes respectively, which form an orthogonal set. Since both modes satisfy $\hat{n} = \hat{u}_\perp \times \hat{u}_\parallel$ and the relationship between electric and magnetic field is:

$$\vec{H}_{TE,TM} = \frac{\hat{n} \times \vec{E}_{TE,TM}}{\eta_0} \quad (16)$$

they are homogeneous plane waves propagating in $\hat{n}$ direction, and their linear combination is able to represent an arbitrary polarization.

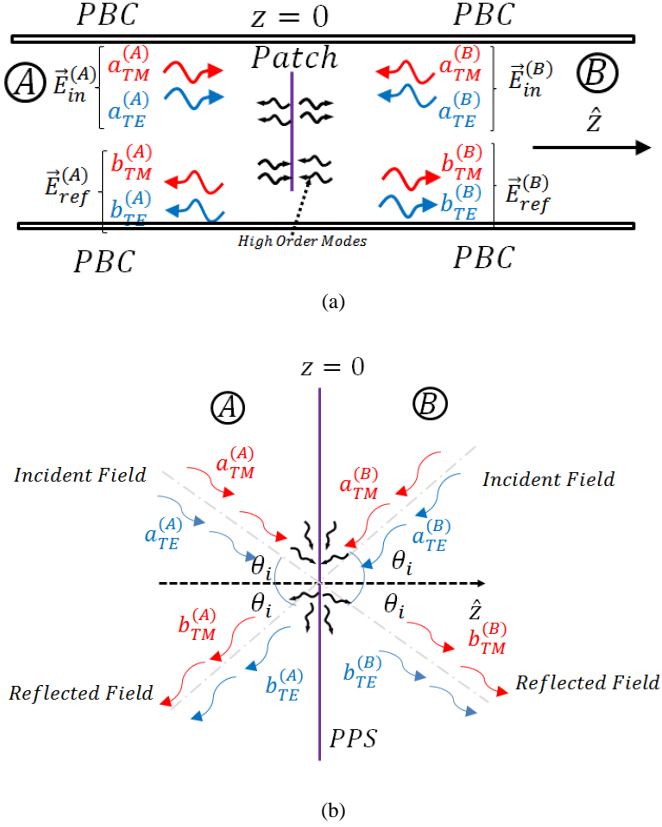

Fig. 2. Scattering of a PPS for oblique incidence within the non-diffraction regime from the perspective of (a) rectangular waveguides with PBC and (b) a free space problem.

Fig. 2 shows the scattering of a PPS under oblique incidence from the perspective of both, a guided system with translational symmetry (Fig. 2a) and a free space problem (Fig. 2b). The modal impedances of the two fundamental modes, according to equations (11) and (12), are:

$$Z_{TE} = \frac{\hat{z} \times \vec{E}_{tTE}}{\vec{H}_{tTE}} = \eta_0 / \cos\theta_i \quad (17)$$

$$Z_{TM} = \frac{\hat{z} \times \vec{E}_{tTM}}{\vec{H}_{tTM}} = \eta_0 \cos\theta_i \quad (18)$$

and the propagation constant, which is equal for both modes, is:

$$k_{z0} = k_{z0(TE)} = k_{z0(TM)} = k_0 \cos\theta_i \quad (19)$$

Let be assumed that the fundamental modes are propagating and the high order modes are evanescent. This scenario is often called the non-diffraction regime or bimode band. In the absence of other sources, if the PPS is excited by a plane wave, combination of TE and TM modes, propagating towards $\hat{n}$, the PPS will generate all the modes at both sides of the discontinuity to fulfill the boundary conditions imposed at the plane $z = 0$ (Fig. 2a). However, the high-order modes are not propagating within the non-diffraction regime because they are considered sufficiently attenuated at a small distance away from the discontinuity. This makes the effects of these modes negligible in the analysis within the bimode band. Under this condition, the incident and reflected fields at both sides of the PPS can be expressed as a combination of the fundamental modes:

$$\begin{aligned}
\vec{E}_{in}^{(A)} &= \left(a_{TE}^{(A)} \cdot \hat{a}_\parallel + a_{TM}^{(A)} \cdot \hat{a}_\perp\right) M(x,y) e^{-jk_{z0}z} \\
\vec{E}_{ref}^{(A)} &= \left(b_{TE}^{(A)} \cdot \hat{a}_\parallel + b_{TM}^{(A)} \cdot \hat{a}_\perp\right) M(x,y) e^{+jk_{z0}z} \\
\vec{E}_{in}^{(B)} &= \left(a_{TE}^{(B)} \cdot \hat{a}_\parallel + a_{TM}^{(B)} \cdot \hat{a}_\perp\right) M(x,y) e^{jk_{z0}z} \\
\vec{E}_{ref}^{(B)} &= \left(b_{TE}^{(B)} \cdot \hat{a}_\parallel + b_{TM}^{(B)} \cdot \hat{a}_\perp\right) M(x,y) e^{-jk_{z0}z}
\end{aligned} \quad (20)$$

being:

$$M(x,y) = e^{-jk_{x0}x} e^{-jk_{y0}y} \quad (21)$$

The complex amplitudes of the incident and reflected fields at $z = 0$ can be related through the tangential fields using the generalized 4-port S matrix. Thus, the difference between the modal impedances of the TE and TM modes is considered in the reference impedances of the 4-port network (Fig. 3).

$$\begin{bmatrix} b_{TE}^{(A)} \\ b_{TM}^{(A)} \\ b_{TE}^{(B)} \\ b_{TM}^{(B)} \end{bmatrix} = \underbrace{\begin{bmatrix} S_{11} & S_{12} & S_{13} & S_{14} \\ S_{21} & S_{22} & S_{23} & S_{24} \\ S_{31} & S_{32} & S_{33} & S_{34} \\ S_{41} & S_{42} & S_{43} & S_{44} \end{bmatrix}}_{Ref:(Z_{TE}, Z_{TM}, Z_{TE}, Z_{TM})} \begin{bmatrix} a_{TE}^{(A)} \\ a_{TM}^{(A)} \\ a_{TE}^{(B)} \\ a_{TM}^{(B)} \end{bmatrix} \quad (22)$$

Assuming that the PEC discontinuity of the PPS is infinitely thin, the orthogonality between modes leads to mode-to-mode continuity at $z = 0$ for an arbitrary pair $(\theta_i, \varphi_i)$, which imposes the following constraints on the S parameters [37]:

$$\begin{aligned}
S_{13} &= 1 + S_{11} \\
S_{24} &= 1 + S_{22} \\
S_{12} &= S_{14}
\end{aligned} \quad (23)$$

Adding the conditions imposed by the input-output symmetry and the reciprocity of the structure, the complete S-matrix will be of the form (24), where the outer subindices indicate the reference impedance of the corresponding port number:

$$S = \begin{bmatrix} S_{11} & S_{12} & 1+S_{11} & S_{12} \\ S_{12} & S_{22} & S_{12} & 1+S_{22} \\ 1+S_{11} & S_{12} & S_{11} & S_{12} \\ S_{12} & 1+S_{22} & S_{12} & S_{22} \end{bmatrix}_{(Z_{TE},Z_{TM},Z_{TE},Z_{TM})} \quad (24)$$

As it can be noted, the scattering of a PPS with arbitrary geometry excited under oblique incidence can be described at each frequency using three independent S parameters within the bimode band, similarly to the case of normal incidence. This fact allows to extend the bimode circuit proposed in [36] to the oblique incidence, provided that each port is referenced to the corresponding modal impedance.

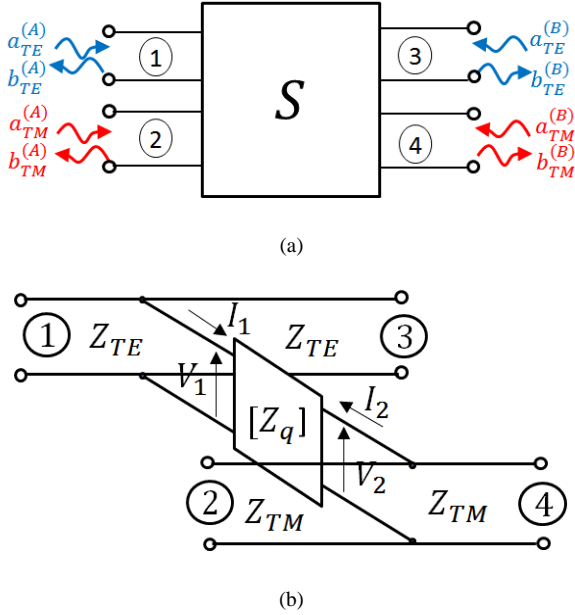

Fig. 3. (a) 4-port network of the PPS modelling the interaction between the fundamental modes for oblique incidence. (b) Equivalent circuit for the PPS discontinuity at oblique incidence, with port numbers and reference impedances corresponding to each mode, and the inner interconnection network.

### B. Bimode equivalent circuit

The above S-matrix restrict the topology of the 4-port network that models accurately the mode interaction phenomena. The impedance matrix of the 4-port network can be derived from the scattering matrix using [40]:

$$S = F(Z - G^H)(Z + G)^{-1}F^{-1}$$
$$Z = F^{-1}(I - S)^{-1}(SG + G^H)F$$
$$F = \text{diag}\left\{1/2\sqrt{|\Re Z_i^{ref}|}\right\}_i \quad (25)$$
$$G = \text{diag}\{Z_i^{ref}\}_i$$

thus obtaining:

$$Z = \begin{bmatrix} [Z_q] & [Z_q] \\ [Z_q] & [Z_q] \end{bmatrix} \quad (26)$$

where

$$[Z_q] = Z[1:2,1:2] = \begin{bmatrix} Z_{11} & Z_{12} \\ Z_{12} & Z_{22} \end{bmatrix} \quad (27)$$

The circuit network that represents the 4-port $Z$ matrix is depicted in Fig. 3b, which is composed by a two-port interconnection network, $[Z_q]$. It must be noted that, since rows and columns are repeated in $Z$, it is singular and thus it has no equivalent admittance matrix $Y$. However, $[Z_q]$ does not suffer from this problem, and it is possible to obtain the admittance matrix of the interconnection network.

Assuming a PPS made up of PEC, i.e., neglecting losses in the conductor, the single-layered structure under study is reciprocal and lossless, and thus, purely reactive. This means that the elements $Z_{11}$ and $Z_{22}$ of $[Z_q]$ are positive real odd functions and they can be expanded in partial fractions (28) with $k_{jj} > 0$. Furthermore, $Z_{12}$ is an odd rational function with simple $j\omega$-axis poles and it can be expressed as (29) where all $k_{12}$ are real, either positive or negative, and satisfying the residue condition (30) [41]. This last condition can be replaced by the short-circuit admittance parameter $Y_{11}$ (31), which is a positive rational real odd function [41].

$$Z_{jj}(s) = k_{jj}^\infty s + \frac{k_{jj}^0}{s} + \sum_{i=1}^{n} \frac{2k_{jj}^i s}{s^2 + \omega_i^2} \quad j = 1,2 \quad (28)$$

$$Z_{12}(s) = k_{12}^\infty s + \frac{k_{12}^0}{s} + \sum_{i=1}^{n} \frac{2k_{12}^i s}{s^2 + \omega_i^2} \quad (29)$$

$$k_{11}^i k_{22}^i - (k_{12}^i)^2 \geq 0 \quad (i = \infty, 0, 1, \ldots n) \quad (30)$$

$$Y_{11} = \frac{Z_{22}}{Z_{11}Z_{22} - Z_{12}^2} \quad (31)$$

The reactance two-port synthesis can be accomplished using the method exposed by Cauer [42], since $[Z_q]$ can be expressed as (32):

$$[Z_q] = \begin{bmatrix} k_{11}^\infty s + \frac{k_{11}^0}{s} + \sum_{i=1}^{n}\frac{2k_{11}^i s}{s^2+\omega_i^2} & k_{12}^\infty s + \frac{k_{12}^0}{s} + \sum_{i=1}^{n}\frac{2k_{12}^i s}{s^2+\omega_i^2} \\ k_{12}^\infty s + \frac{k_{12}^0}{s} + \sum_{i=1}^{n}\frac{2k_{12}^i s}{s^2+\omega_i^2} & k_{22}^\infty s + \frac{k_{22}^0}{s} + \sum_{i=1}^{n}\frac{2k_{22}^i s}{s^2+\omega_i^2} \end{bmatrix} \quad (32)$$

Given that this method uses the partial fraction decomposition of the $Z_{ij}$ elements, it can be regarded as a generalization of Foster's theorem [43]. Therefore, it is possible to expand $[Z_q]$ as a sum of simple terms following Foster's reactance synthesis (33).

$$[Z_q] = [Z^\infty] + [Z^0] + \sum_{i=1}^{n}[Z^i] =$$

$$= \begin{bmatrix} k_{11}^\infty s & k_{12}^\infty s \\ k_{12}^\infty s & k_{22}^\infty s \end{bmatrix} + \begin{bmatrix} \dfrac{k_{11}^0}{s} & \dfrac{k_{12}^0}{s} \\ \dfrac{k_{12}^0}{s} & \dfrac{k_{22}^0}{s} \end{bmatrix}$$

$$+ \sum_{i=1}^{n} \begin{bmatrix} \dfrac{2k_{11}^i s}{s^2+\omega_i^2} & \dfrac{2k_{12}^i s}{s^2+\omega_i^2} \\ \dfrac{2k_{12}^i s}{s^2+\omega_i^2} & \dfrac{2k_{22}^i s}{s^2+\omega_i^2} \end{bmatrix} \quad (33)$$

Fig. 4 shows the Cauer's circuit with the expansion of matrix $[Z_q]$. In this circuit, all the L's and C's elements obtained by the expansion of $Z_{11}$, $Z_{22}$ and $Z_{12}$ are real and positive, while the ideal transformers $(1:a^i)$, $(i = \infty, 0, 1, ..n)$ (Fig. 4) are positive or negative according to the sign of $k_{12}^i$ in order to verify the residue condition (30). In conclusion, all the matrices $[Z^\infty], [Z^0]$ and $\sum_{i=1}^{n}[Z^i]$ in (33) are realizable by the elementary two-ports in Fig. 4

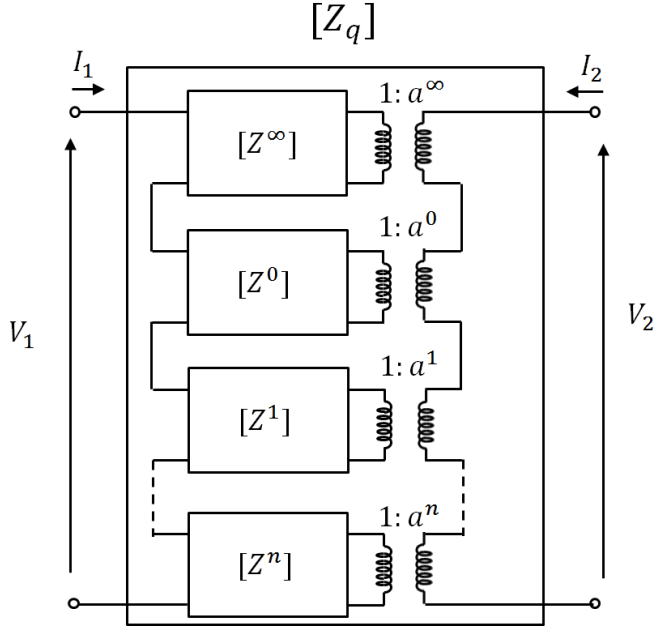

Fig. 4. Cauer circuit modeled for the synthesis of the two-port network $[Z_q]$.

Therefore, an arbitrary PPS under oblique incidence may be accurately described by a network composed only by frequency-independent, lumped LC elements and ideal transformers. However, once the circuit realizability of the structure under study has been proved, T and π two-port networks (Fig. 5) will be considered for the sake of functionality in order to work with a more suitable circuit than that proposed by Cauer. The relationship between the branch immittances and the open-circuit and short-circuit parameters are shown in (34) and (35).

$$\text{T network} \begin{cases} Z_a = Z_{11} - Z_{12} \\ Z_b = Z_{12} \\ Z_c = Z_{22} - Z_{12} \end{cases} \quad (34)$$

$$\pi \text{ network} \begin{cases} Y_a = Y_{11} + Y_{12} \\ Y_b = -Y_{12} \\ Y_c = Y_{22} + Y_{12} \end{cases} \quad (35)$$

Since the immittances $Z_a$, $Z_b$, $Z_c$, $Y_a$, $Y_b$, $Y_c$ are not necessarily positive real odd functions, negative values of L and C are obtained in the examples presented in Section III. It is also clear that the circuit is not constrained to a single resonator; it can accommodate any needed number of resonances within the bimode band, represented by the sum in (32) and (33), from one (or none) to $n$ resonators. This number is strongly dependent upon the geometry of the cell.

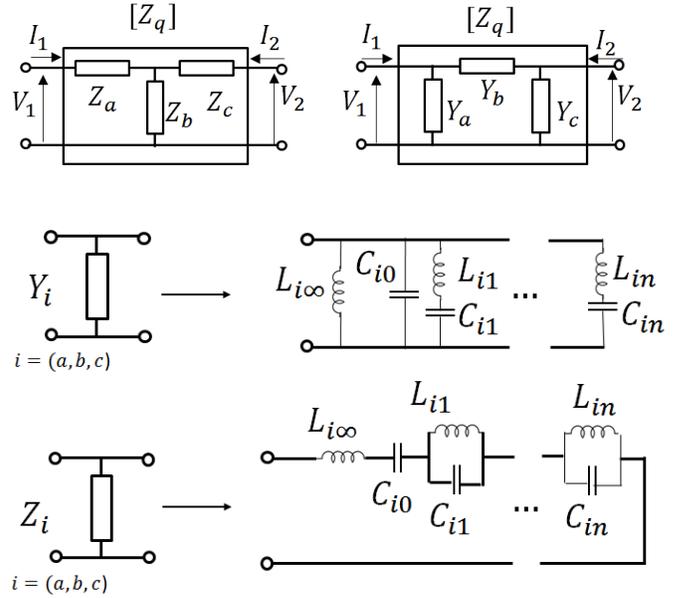

Fig. 5. π- and T-type topology for modeling the interconnection network [Zq].

Once the lumped elements for an arbitrary PPS and angle of incidence are obtained using the corresponding EM simulation, a multilayer structure made up of several PPSs and spacers can be easily computed by only cascading the 4-port circuits that represent each PPS connected by transmission lines whose impedances are $Z_{TE}$ and $Z_{TM}$. As aforementioned, the proposed circuit is able to represent the electrical behavior of the discontinuity within the bimode band. For instance, according to (7), a square period (the most common case) exhibits four pairs of degenerated high order modes, for $(|m| = 1, n = 0)$ and $(m = 0, |n| = 1)$. Since the cutoff frequency of each one depends on the period of the PPS and the direction of propagation of the field, given by $\theta_i$ and $\varphi_i$, the minimum value of these cutoff frequencies will determine the range of validity of the circuit.

### C. Further Generalizations for Dielectrics

The design process of advanced PPS-based devices may need the introduction of more advanced features such as the use of dielectric spacers to fulfill the required specifications. To cope with them, the modal impedances and the propagation constant must be modified in the equivalent circuit, since (23) is not satisfied.

If a PPS is placed between vacuum and a dielectric of relative permittivity $\varepsilon_{r1}$ and thickness $d$ (see Fig. 6), the modes in the dielectric must fulfill the Helmholtz equation (assuming $k = k_0\sqrt{\varepsilon_{r1}}$) and the periodic conditions given by (2), which leads to:

$$k_{z0}^{\varepsilon_1} = \sqrt{\varepsilon_{r1}k_0^2 - (k_{x0}^2 + k_{y0}^2)} \quad (36)$$

thus, the fundamental modes are plane waves propagating in the dielectric at the direction:

$$\hat{n} = \frac{\sin\theta_i \cos\varphi_i \cdot \hat{x} + \sin\theta_i \sin\varphi_i \cdot \hat{y}}{\sqrt{\varepsilon_{r1}}} + \sqrt{1 - \frac{\sin^2\theta_i}{\varepsilon_{r1}}} \cdot \hat{z} \quad (37)$$

Therefore, the mode impedances and the propagation constant are:

$$Z_{TE}^{\varepsilon_1} = \frac{\omega\mu_0}{k_z} = \frac{\eta_0}{\sqrt{\varepsilon_{r1} - \sin^2(\theta_i)}} \quad (38)$$

$$Z_{TM}^{\varepsilon_1} = \frac{k_z}{\omega\varepsilon_1} = \frac{\eta_0}{\varepsilon_{r1}}\sqrt{\varepsilon_{r1} - \sin^2(\theta_i)} \quad (39)$$

$$k_{z0}^{\varepsilon_1} = k_0\sqrt{\varepsilon_{r1} - \sin^2(\theta_i)} \quad (40)$$

Under these conditions, the values of the lumped elements comprising the interconnection network $[Z_{q1}]$ must include the effect of the dielectric. The circuit model showed in Fig. 6 represents an approach with increasing accuracy as the distance $d$ is higher. When $d$ tends to infinity, i.e., the final structure consists of two indefinite media the capacitances of the interconnection network $C_i'$ will be:

$$C_i' = C_i \cdot \overline{\varepsilon_r} = C_i \cdot \frac{\varepsilon_{r1} + \varepsilon_{r0}}{2} \quad (41)$$

with $C_i$ the capacitances obtained in the case of vacuum. The new inductances $L_i'$ are the same as those obtained for the vacuum, $L_i$, since the stored magnetic energy does not change [37]. However, if the dielectric exhibits a finite thickness $d$, the approach of (41) must be taken carefully. The exact values of the capacitances and inductances could be extracted from an EM simulation that considers the presence of the dielectric. However, the approach of indefinite media may be usually valid for thicknesses thinner than $\lambda/10$, as demonstrated in the next section.

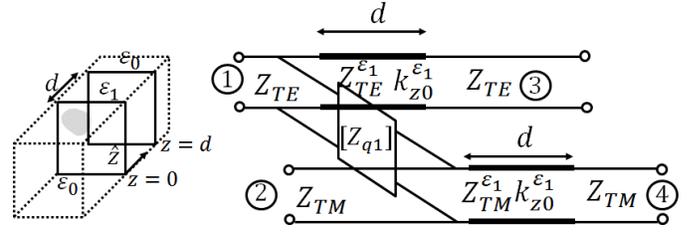

Fig. 6. Multilayer cell composed of two PPS of arbitrary geometry separated by a dielectric spacer of permittivity $\varepsilon_1 = \varepsilon_{r1}\varepsilon_0$ and thickness $d$. Bimode equivalent circuit of the complete structure for oblique incidence.

Once the values of the lumped elements consider the effect of the dielectrics, it is direct to simulate a multilayer structure made up of a cascade of PPSs and dielectrics using for example the T parameters of each element [44].

*D. Advantages and limitations of the proposed equivalent circuit.*

All the techniques and improvements presented in the previous subsections make the equivalent circuit advantageous against other approaches like phenomenological circuits or brute-force optimization. The advantages of the proposed circuit can be summarized as follows:

1) The topology can be selected beforehand and is not dependent on any physical shape or constraint. Moreover, the number of circuit elements is minimum, as all the energy interactions in the structure under analysis are encompassed in the lumped elements, unlike the phenomenological circuits.

2) The circuit is exclusively made up of frequency-independent lumped elements. It fulfills the Foster's theorem, integrating the systematization in the calculation of the elements for an arbitrary geometry and an improving the memory usage since the frequency behavior must not be stored. Moreover, the theorem implies that the zero-pole scheme of $Z_{ij}$ (or, equally $Z_a, Z_b, Z_c$) is the same for all of them. In this way, the extraction of each resonator implies two parameters: the pole and the residue.

3) Some properties can be used to obtain circuit elements of a structure without computing its response using an EM simulator, by means of the scale factor $F$. If the lumped elements, capacitors $C_k$ and coils $L_k$ of a physical structure with dimensions $D_1, D_2 \ldots D_n$ are obtained for an angle of incidence, the equivalent circuit that results from scaling $D_1, D_2 \ldots D_n$ by a multiplicative factor $F$, while keeping the angle of incidence constant, is the same as before but dividing the $L_k$ and $C_k$ values by $F$. In addition, it is possible to model the introduction of a dielectric layer just by scaling only the capacitances $C_k$ without a new EM simulation.

4) Properties 2) and 3) allow obtaining the circuit values of a PPS as a function of its geometric parameters (parameterization) by only simulating the cell under vacuum conditions at both sides of the discontinuity for a coarse grid of parameters. Once they are extracted from the EM simulations, the new lumped elements for arbitrary dielectrics and/or scaled dimensions can be easily computed. It also allows obtaining the circuit values for geometric parameters within the coarse grid without further EM simulation, just by interpolating the previously extracted values. Moreover, once the PPS (2D problem) is parameterized, the simulation of a stacked cell comprising an arbitrary number of PPSs at any frequency can be performed without computational effort (each circuit simulation takes microseconds) by cascading the circuits, avoiding the large compute time of the EM simulation. This is especially advantageous if the structure is made up of a large number of layers.

5) The most important property of the proposed circuit is the direct use of the advance synthesis techniques from the circuit theory. This will allow the design of PPS-based devices in oblique incidence without resorting to costly full-wave optimizations or, at least, reducing dramatically the computational effort. Moreover, it will provide new insights to the nature of a geometry by inspecting its circuit elements, allowing *a priori* knowledge about the behavior of the structure. It must be noted that the equivalent circuit changes its port impedances for each pair $(\theta, \varphi)$ and consequently for obtaining a complete space of angular samples, the angle of incidence must be discretized too. Nonetheless, this is not a severe problem because the angular information is computed once and then stored, thereby reducing the simulating burden when a new device is designed.

The proposed equivalent circuit presents some limitations:

a) Since it is limited to lossless dielectrics, a final adjustment by EM simulation will be necessary, or at least checking if they are not significant.

b) The distance between different layers must be large enough to neglect the contribution of the non-propagating higher modes. Without this assumption, these modes will modify the response within the bimode band, that cannot be represented by an equivalent circuit. This fact makes the circuit suitable to simulate structures comprising dielectrics thicker than $\lambda/10$ (where $\lambda$ is the wavelength in the dielectric), although thinner dielectrics could be considered if the structure is made of dielectric-air interfaces [36]. An example will be presented in the next section to point out this issue.

c) The circuit is not able to represent the behavior of high order modes when they are propagating. Note that a large number of devices in the literature are designed to operate under the influence of a large number of propagating modes, especially those exhibiting very thin dielectrics [15], although at the expense of using pure EM optimizations.

### III. STRUCTURES SIMULATED BY THE EQUIVALENT CIRCUIT

In order to validate the proposed equivalent circuit, different unit cells used to develop PPS devices (single layer and multilayer) are evaluated including the effects of the dielectrics, comparing the results obtained using an electromagnetic (EM) simulator [45]. and the bimode Foster's circuit.

#### A. Example 1: Rotated dipole in vacuum and square period

As shown in Fig. 7a, the first PPS analyzed is a single layer cell based on a rotated dipole under oblique incidence. It has been used previously to design PPS devices to rotate polarization or control overall XP of planar antennas [46].

As a first approximation to the analysis, a single parameterization has been studied for a square period ($W = 0.5$ mm, $\alpha = 81°$, $L = 9$ mm, $P = P_x = P_y = 10$ mm). The angles of incidence were fixed to $\theta_i = 20°$, $\varphi_i = 30°$. The arbitrary nature of the geometry (no symmetries for the general case, neither limitation regarding the angle of incidence) constraints the interconnection topology chosen to one that appropriately models the energy exchange between the fundamental modes; hence, a Π topology (Fig. 7b) was chosen. The S parameters of the structure were calculated using the EM simulator for the bimode band, and the admittances that compose the network ($Y_a$, $Y_b$, $Y_c$) were extracted. Since there is a resonance within the band (Fig. 8a), they must be modelled by means of a series LC resonator (if developed in the second Foster form). However, the behavior at the far limits of the band is more accurately represented using an extra parallel capacitor. A single curve fitting tool was used to extract the elements that compose each admittance. The extracted lumped elements according to Fig. 7 are summarized in Table I.

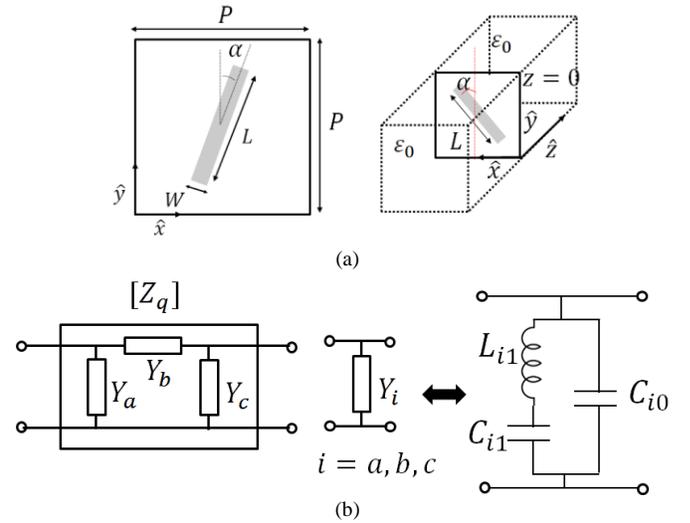

Fig. 7. (a) View of the rotated dipole unit cell with its geometric parameters and waveguide viewpoint. (b) Interconnection network as a Π topology with inner admittances composed of a series LC resonator with a shunt capacitor.

Fig. 8b and Fig. 8c show total agreement in both amplitude and phase between the S parameters calculated by the EM simulator (CST) and those calculated from the equivalent circuit within the bimode band. Regarding this point the cutoff frequency of the first high order mode is 23.4 GHz. Therefore, this circuit is able to model accurately the rotated dipole under oblique incidence. Assuming that the smooth variation of some of the dimensional parameters will not change the fundamental nature of the unit cell, a whole range of geometries can be modelled with this topology.

TABLE I
CIRCUIT AND PHYSICAL PARAMETERS OF THE PPS ANALYZED IN THE EXAMPLE 1

| Circuit Parameter | Value |
|---|---|
| $C_{a0}, C_{b0}, C_{c0}$ (fF) | -0.2826, 0.6998, 1.2905 |
| $C_{a1}, C_{b1}, C_{c1}$ (fF) | -2.6069, 4.2376, 6.7745 |
| $L_{a1}, L_{b1}, L_{c1}$ (nH) | -37.8872, 23.3017, 14.5758 |
| **Physical Parameter** | **Value** |
| $P = P_{x,y}, L, W$ (mm) | 10, 9, 0.5 |
| $\alpha$ (º) | 81 |
| Angle of incidence: $\theta_i$ (º), $\varphi_i$ (º) | 20, 30 |

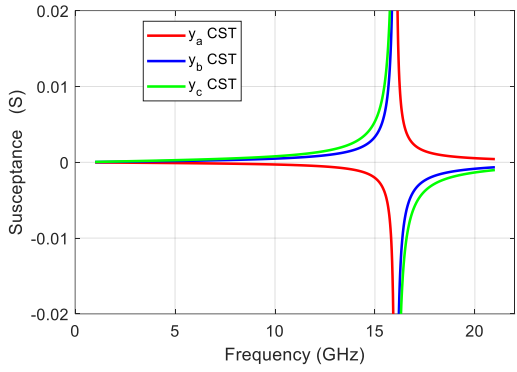
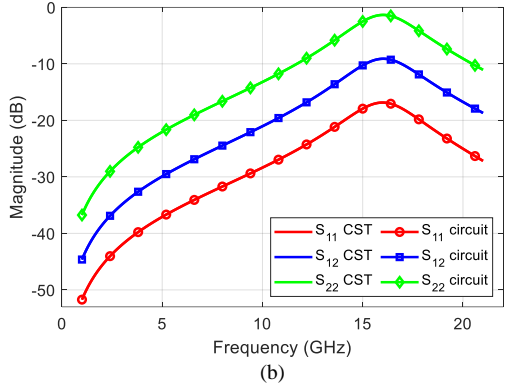
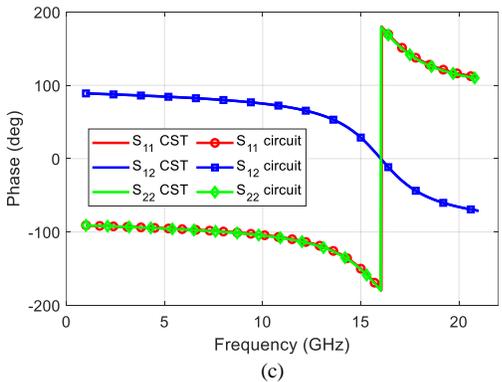

Fig. 8. Simulation of a rotated dipole with the circuit and physical parameters of Table I. (a) Susceptances of the Π equivalent. (b) and (c) Comparison between S parameters from EM simulation (CST) and the equivalent circuit, in magnitude and phase.

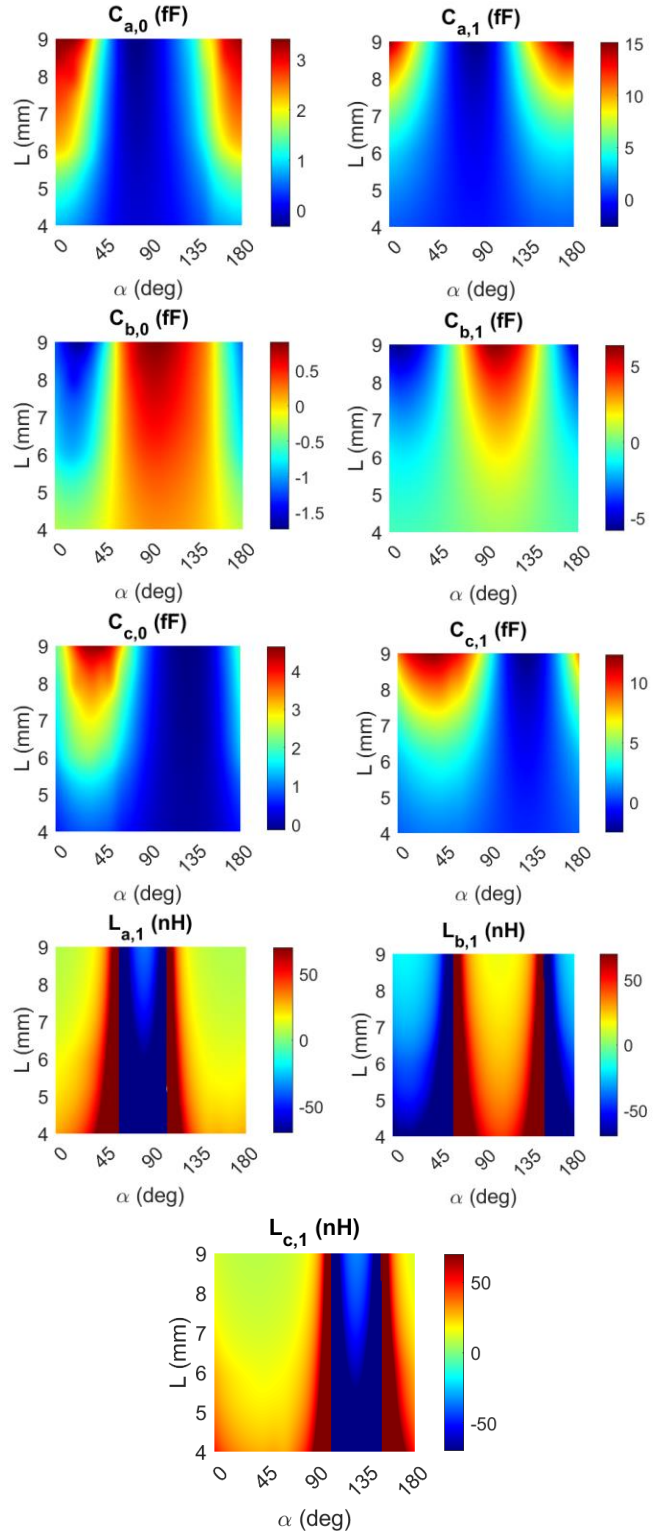

Fig. 9. Maps of discrete capacitances and inductances of the Π equivalent circuit (Fig. 7b) of the rotated dipole across the parametric sweep (L and $\alpha$).

Fixing the period $P$ and the width of the dipole $W$, a parameter sweep was performed with parameters $L$ and $\alpha$. The discrete lumped elements were extracted for each simulation, and a bidimensional map of inductances and capacitances was constructed (Fig. 9). The validity of these results was checked by calculating the $S_{11}$ of the whole sweep at 10 GHz using the equivalent circuits and comparing them to the results obtained by the EM simulation. The results from the circuit accurately match, both in amplitude and phase the simulations Fig. (10).

Note that, whereas in [36] the lumped elements from the geometries with $\alpha \in (45°, 90°)$ and $\alpha \in (90°, 180°)$ could be easily computed from the values in $\alpha \in [0°, 45°]$, this does not apply to oblique incidence, since the symmetry between mode fields no longer exists, and the reference impedances are different.

As seen in the previous section, it is important to note that the period and the angle of incidence remain constant across simulations. Otherwise, it would affect the reference impedances used in the ports, and the validity of the interconnection network would not be guaranteed.

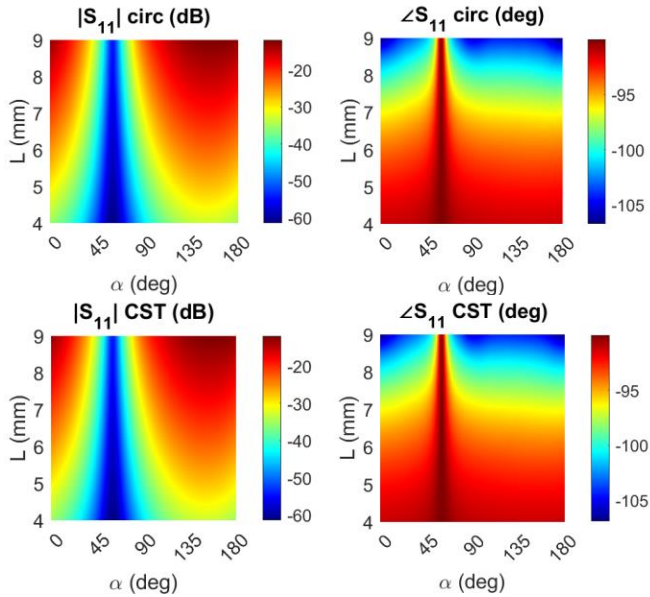

Fig. 10. $S_{11}$ from parametric sweep at 10 GHz, as calculated by the EM simulator (CST) and calculated from the circuit, in magnitude and phase.

### B. Example 2: Stacked Rotated Dipoles

The next structure to analyze is a multilayer configuration filled with a dielectric between each PPS as shown in Fig. 11. The dielectric between layers has a relative permittivity of $\varepsilon_{r1} = 3$ and thickness $d = 10$ mm, which effectively reduces the cutoff frequency to 14.8 GHz and limits the validity of the equivalent circuit.

The interconnection networks, represented in Fig. 11b by $[Z_{q_1}]$ and $[Z_{q_2}]$, are obtained from the maps extracted in the parametric sweep for vacuum environment depicted in Fig. 9, by applying the factor $F$ explained in the previous section regarding the dielectrics. It must be remembered that the only modification that should be performed to the equivalent circuit in this case is to multiply the capacitors obtained from the cell under vacuum by the corresponding factor, providing that the dielectric thickness is thick enough to assume the effective permittivity of the indefinite media. Otherwise, further simulations should be carried out. The physical and circuit parameters according to Fig. 11 are disclosed in Table II, using a Π-type equivalent for both the first and second layer, and the same lumped elements than in Fig. 7b.

The whole structure has been simulated using an EM simulator, and the results compared to those obtained using the cascade connection of the two equivalent circuits and the transmission line. The results of the circuit are in total agreement with the full-wave simulation in terms of S parameters as depicted in Fig. 12, which validates the modelling of dielectrics and multilayer devices using the equivalent circuit approach.

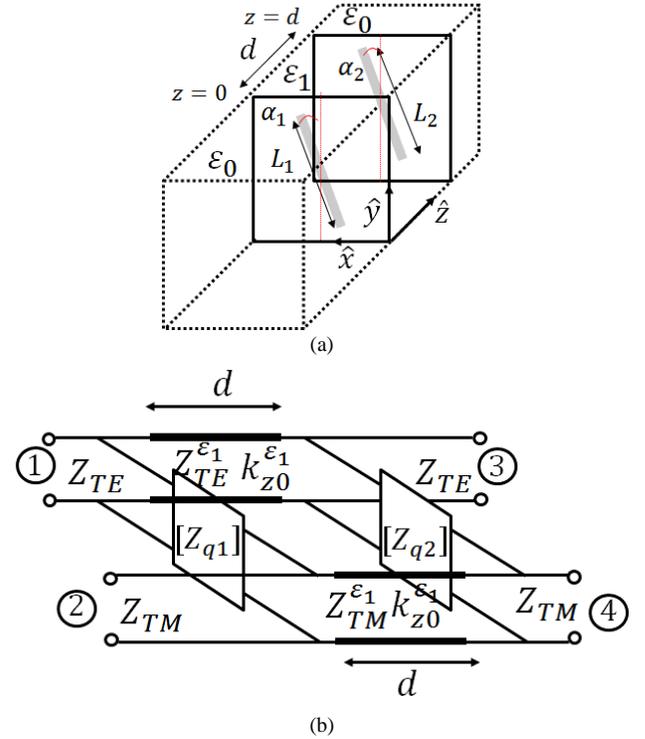

Fig. 11. Two stacked dipoles separated by a dielectric of permittivity $\varepsilon_1 = \varepsilon_0 \varepsilon_{r1}$ (a) 3-D geometry and (b) Equivalent circuit.

TABLE II
PARAMETERS OF THE CELL ANALYZED IN THE EXAMPLE 2 ($\varepsilon_{r1} = 3, d = 10\ mm$)

| Circuit Parameter | Layer 1 $[Z_{q1}]$ | Layer 2 $[Z_{q2}]$ |
|---|---|---|
| $C_{a0}, C_{b0}, C_{c0}$ (fF) | 2.9640, -1.6616, 4.4834 | 2.70183, 0.2999, 0.0470 |
| $C_{a1}, C_{b1}, C_{c1}$ (fF) | 7.8184, -4.7368, 12.0196 | 18.489, 2.1587, -1.9330 |
| $L_{a1}, L_{b1}, L_{c1}$ (nH) | 16.943, -27.965, 11.0210 | 8.8781, 76.038, -84.916 |
| **Physical Parameter** | **Layer 1** | **Layer 2** |
| $P_{x,y} = P$ (mm) | 10 | 10 |
| $L, W$ (mm) | 7, 0.5 | 8.5, 0.5 |
| $\alpha$ (°) | 27 | 144 |
| $\theta_i$ (°), $\varphi_i$ (°) | 20, 30 | 20, 30 |

Fig. 13 represents a comparison between the $S_{11}$ of both EM and circuit simulations for several values of the transmission length $d$. In the circuit model, the lumped elements for all the thicknesses are the same and corresponded to the approach of indefinite media. As it can be seen, the circuit is able to simulate the electrical behavior of the cell accurately even when a thin dielectric like $d = 3$ mm $\approx \lambda/10$ at central frequency, is used. Moreover, it can be also appreciated that the approach of indefinite media is suitable in this case. It must be highlighted that the effects of the evanescent modes on the circuit depend not only on the thickness between layers, but on the particular geometry of the PPS too, that fixes the mode magnitudes.

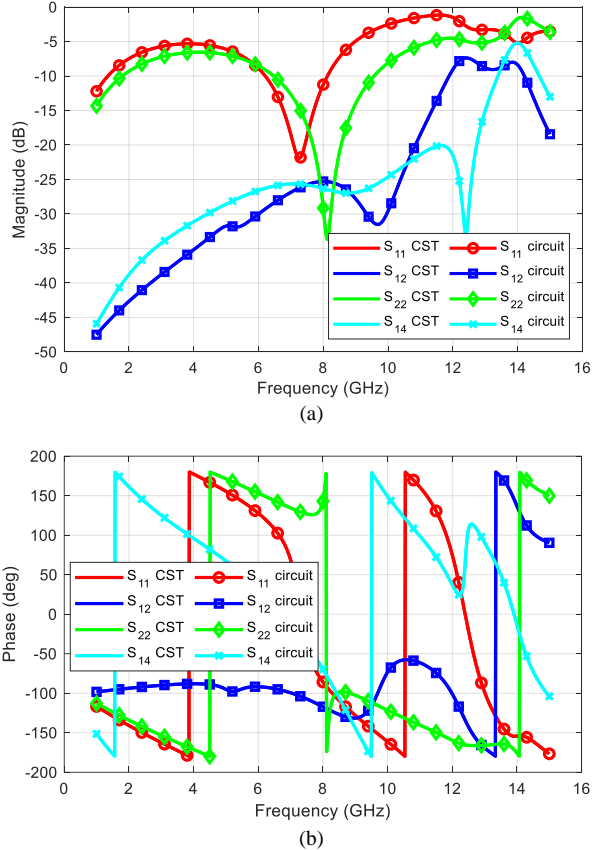

Fig. 12. Comparison between the S parameters of two stacked dipoles separated by a dielectric according to the dimensions in Table II, as obtained through a full wave simulation and using the equivalent circuit. (a) Magnitude (b) Phase.

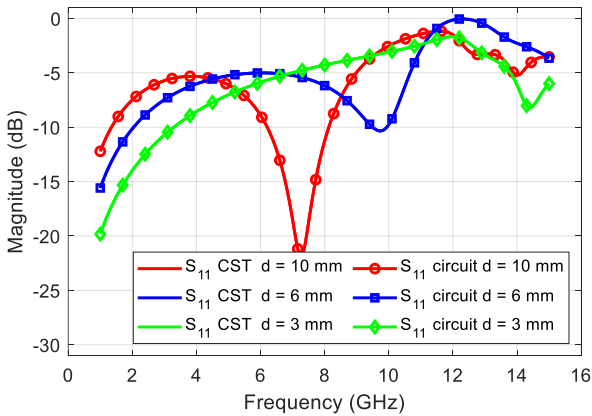

Fig. 13. Comparison of the magnitude of the $S_{11}$ parameter of two stacked dipoles, as obtained through a full wave simulation (solid) and using the equivalent circuit and interpolation (dashed), for various dielectric thicknesses.

## IV. POLARIZATION ROTATOR DESIGN WITH EXPERIMENTAL VALIDATION

In this section the design process of a reflective 90º polarization rotator, its manufacturing and testing is presented. It is a clear demonstration of the accuracy of the equivalent circuit and its potential for designing polarization control devices.

### A. Single-Layer, Reflective 90º Polarization Rotator

The translation of the polarization requirements working in oblique incidence between TE and TM modes can be simplified in the case where the electric field is totally contained in one of the modes, for example, the TE mode. In this case, a polarization rotator is simply a device that totally exchanges energy from one mode to the other. For this purpose, the geometry selected is the double-L cell shown in Fig. 14a, which has proven valuable for other polarization control devices [4]. Thus, the complete structure of the reflective rotator is shown in Fig. 14a, composed of the PPS, the supporting dielectric and the ground plane.

A reflective rotator working at 25 GHz under the angle of incidence $\theta_i = 30º$, $\varphi_i = 0º$ is designed using the proposed equivalent circuit. Given the general guidelines introduced in [4], the period of the PPS must be as small as possible to support the required resonances in the non-diffraction band, while diminishing the interaction of higher-order modes that may propagate through the substrate. That is reason to choose a period of $P = 3.5$ mm ($\lambda/2$ at 25 GHz in the dielectric). The substrate for manufacturing the device has a thickness of $d = 0.8$ mm and a relative permittivity of $\varepsilon_r = 3.2$. As explained in Section II, substrate losses are not considered in the equivalent circuit. Taking into account the manufacturing process constraints, i.e., the width of the arms $W$ and the separation between arms $S$ higher than 0.2 mm, there is a limitation on the range of possible geometries, easing to perform a parameter sweep over the geometric variables.

It is found that the most suitable topology that accurately models the geometry is a T-type two-port network, where each impedance is expanded (first Foster form) as a shunt LC resonator in series with a capacitor as shown in Fig. 14b. The lumped elements according to Fig. 14b are extracted for a parameter sweep of the geometrical variables $W$, $L$ and $S$ at the angle of incidence $\theta_i = 30º$, $\varphi_i = 0º$. Five samples for each variable are considered, in total 125 EM-simulations, which are enough to interpolate the other values. The lumped elements are calculated under vacuum environment, and the corrected elements to consider the dielectric are again obtained with the approach of infinite medium.

It is direct to translate the PPS structure, the dielectric support, and the ground plane to a complete equivalent circuit shown in Fig. 14c. The existence of this circuit allows also to incorporate the system specifications to the equivalent circuit properties. Particularly, the condition that all energy must be transmitted from the TE mode to the TM mode means that when the circuit is excited in the TE port, referenced by impedance $Z_{TE}$, all the power must be absorbed by the circuit, i.e., there is impedance matching at port 1, and the input admittance $Y_{in}$ fulfills:

$$Y_{in} = 1/Z_{TE}^* \quad (42)$$

Since all the components of the circuit are known, the input admittance can be expressed in terms of $Y_{LE}$ and $Y_{LM}$, the impedances seen at the other side of the PPS discontinuity which correspond to transmission lines terminated in short-circuit, and $Z_{TM}$, the modal impedance of the TM mode. Therefore, the input admittance $Y_{in}$ can be expressed as:

$$Y_{in} = Y_{LE} + Y_{11} - \frac{Y_{12} \cdot Y_{21}}{(Y_{LM} + 1/Z_{TM}) + Y_{22}} \quad (43)$$

Here $Y_{ij}$ are the entries of the admittance matrix $[Y_q] = [Z_q]^{-1}$ of the interconnection network, which are considered to simplify the expression above.

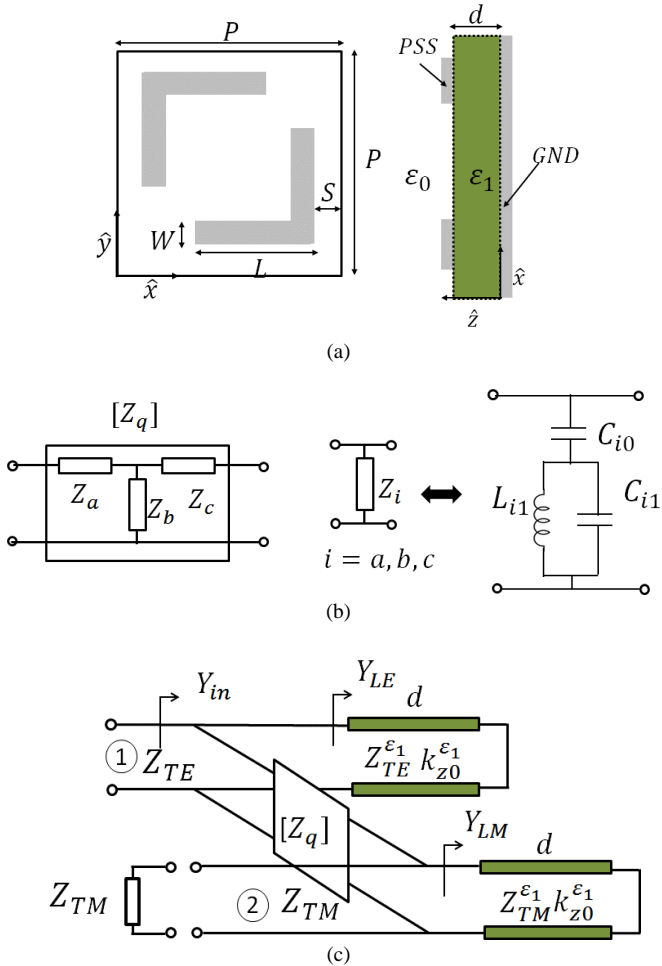

(a)

(b)

(c)

Fig. 14. (a) View of a double-L unit cell with its geometric parameters and side view of the structure. (b) Interconnection network as a T topology, with inner impedances composed of a shunt LC resonator in series with a capacitor, (c) Equivalent circuit of the complete structure.

At this moment since the design equations are known, any optimization technique can be used to find the circuit that best satisfies the equation (42). The developed procedure creates a function that maps the geometric variables to the lumped element, and then evaluates the resulting circuit at the desired frequency (25 GHz). In this way the parameters forming the desired input admittance are found. The computing time is in the order of seconds since there is no EM simulations of the structure, in the process. The advantage is double; on the one hand the time design is dramatically reduce, on the other the existence of a geometry is guaranteed in advance.

The initial design following the above procedure leads to these dimensions: $L$ = 1.85 mm, $S$ = 0.35 mm and $W$ = 0.25 mm. As previously explained, the values given by the equivalent circuit approach is the initial point for fine-tuning with the EM simulator. The final dimensions are collected in Table III, and as it can be observed are quite close to those obtained by the equivalent circuit.

TABLE III
CIRCUIT AND PHYSICAL PARAMETERS OF THE POLARIZATION ROTATOR

| Circuit Parameter | Value |
|---|---|
| $C_{a0}, C_{b0}, C_{c0}$ (fF) | 4.3093, -11.8713, 4.2909 |
| $C_{a1}, C_{b1}, C_{c1}$ (fF) | 8.7758, 8.6701, 5.7221 |
| $L_{a1}, L_{a2}, L_{a3}$ (nH) | 0.4350, 0.4404, 0.6672 |
| **Physical Parameter** | **Value** |
| $P, d$ (mm) | 3.5, 0.8 |
| $L, S, W$ (mm) | 1.86, 0.375, 0.28 |
| Angle of incidence: $\theta_i, \varphi_i$ (°) | 30, 0 |

### B. Polarization rotator, manufacturing and experimental characterization

In order to verify the equivalent circuit as a suitable design tool the polarization rotator has been manufactured and measured. The GML1000 substrate has the following characteristic: permittivity $\varepsilon_r$ = 3.2, $\tan \delta$ = 0.003, and thickness $d$ = 0.762 mm. On the upper copper layer, the shapes have been etched accordingly to the dimensions. The dimensions of the manufactured prototype are 112x112 mm, with a 32x32 square array printed as shown in Fig. 15. This size is big enough to achieve a very low illumination level (-13 dB) by the feed horns on the edges of the periodic surface compared to the center level. The field that impinges on the surface can be considered plane enough to be treated as a plane wave under the selected angle of incidence ($\theta_i$ = 30°, $\varphi_i$ = 0°). As explained before, exciting the TE mode only means that linear polarization along the $\hat{y}$ axis is needed (Fig. 15). This was achieved aligning the axis of the feeder to that of the PPS on the stand. The receiver horn takes two possible positions: one, with the same axis as the transmitter, to measure the XP component $S_{11}$ and one rotated 90° to measure the insertion losses $S_{12}$.

Fig. 16 shows the comparison between the EM simulation, the response of the equivalent circuit with elements values collected in Table III and the measurements. It can be observed that the circuit predicts the minimum in XP reflection with a small frequency shift. This is due to the coupling effect from higher-order modes because of the substrate thickness ($d$ = 0.8 mm = $\lambda/9$). However, the bimode equivalent circuit is suitable to design in this case even if high order modes are propagating.

Fig. 16 also shows the results from the measurements, which demonstrate that the manufactured device has achieved an XP level of -30 dB at central frequency 25 GHz, and a bandwidth of 2 GHz (8%) at -10 dB.

There is a very good match between the full-wave and the measurements. This indicates that the proposed circuit is valid to design polarization converters at oblique incidence and opens up the possibility to design more complex devices exhibiting several layers and/or different periods.

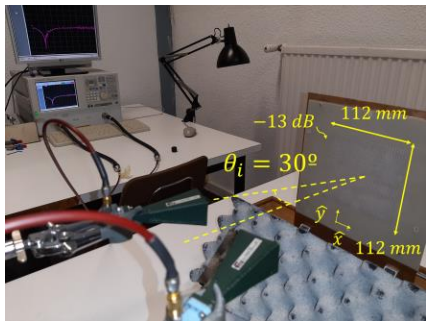

Fig. 15. Setup for measuring the designed 90-rotator at oblique incidence. The horns are placed to measure the XP level $S_{11}$.

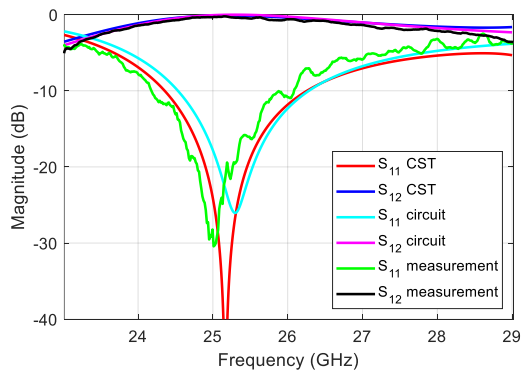

Fig. 16. Comparison between the EM simulation, the equivalent circuit and the measurements of the reflective polarization rotator.

## V. CONCLUSION

A bimode Foster equivalent circuit for arbitrary geometries under oblique incidence has been presented. The proposed circuit is able to model a wide array of elements without changing its topology and keeping the discrete elements independent of frequency. It is also able to properly model multilayer structures with dielectrics, as long as the space between layers is enough to neglect the effect of higher-order modes.

The proposed circuit has been validated by comparing with the results obtained by a full-wave simulator for the case of a rotated dipole. It was also tested against a multilayer structure composed of two stacked rotated dipoles with dielectric inside. In both cases, the equivalent circuit was able to predict the behavior of the structure and to interpolate new geometries not previously full-wave simulated.

The circuit was also used to design a reflexive 90º polarization rotator. The results obtained with a direct search using the equivalent circuit were used as a starting point for a final EM optimization, which offered the final dimensions of the structure. The device was manufactured and tested, achieving an XP level better than -30 dB at the frequency of operation, and showing good agreement with the EM simulation and the equivalent circuit.